\documentclass[aps,prb,twocolumn]{revtex4}

\usepackage{graphicx}
\usepackage{dcolumn}
\usepackage{bm}
\usepackage{longtable}

\begin{document}



\title{First-principles study on 
the intermediate compounds of LiBH$_4$}

\author{Nobuko Ohba, Kazutoshi Miwa, Masakazu Aoki, Tatsuo Noritake, and Shin-ichi Towata}
\affiliation{ 
Toyota Central Research $\& $ Development Laboratories, Inc.,  
Nagakute, Aichi 480-1192, Japan }
\author{Yuko Nakamori, and Shin-ichi Orimo}
\affiliation{
Institute for Materials Research, Tohoku University, Sendai 980-8577, Japan}
\author{Andreas Z{\"u}ttel}
\affiliation{
Physics Department, University of Fribourg, Perolles, CH-1700 Fribourg, Switzerland}

\date{\today}

\begin{abstract}
We report the results of the first-principles calculation 
on the intermediate compounds of LiBH$_4$. 
The stability 
of LiB$_3$H$_8$ and Li$_2$B$_n$H$_n (n=5-12)$ 
has been examined 
with the ultrasoft pseudopotential method based on the density 
functional theory. 
Theoretical prediction has suggested that monoclinic Li$_2$B$_{12}$H$_{12}$ 
is the most stable among the candidate materials.  
We propose the following hydriding/dehydriding process of LiBH$_4$ via 
this intermediate compound : 
LiBH$_4 \leftrightarrow  \frac{1}{12}$Li$_{2}$B$_{12}$H$_{12} + \frac{5}{6}$ 
LiH $+ \frac{13}{12}$H$_2 \leftrightarrow $LiH $+$ B $+ \frac{3}{2} $H$_2$. 
The hydrogen content and enthalpy of the first reaction 
are estimated to be $10$~mass\% and $56$~kJ/mol~H$_2$, respectively, 
and those of the second reaction are $4$~mass\% and $125$~kJ/mol~H$_2$. 
They are in good agreement with experimental results 
of the thermal desorption spectra of LiBH$_4$. 
Our calculation has predicted 
that the bending modes for the $\Gamma$-phonon frequencies 
of monoclinic Li$_2$B$_{12}$H$_{12}$ are lower 
than that of LiBH$_4$, while stretching modes are higher. 
These results are very useful for the experimental
search and identification of possible intermediate compounds. 


\end{abstract}

\maketitle

\section{Introduction} 
\label{sec:int}
Hydrogen is the most promising environmentally 
clean energy carrier to replace fossil fuels. 
The use of hydrogen-based energy in practical applications 
such as fuel cell vehicles, however, 
requires the development of safe and efficient hydrogen storage technology. 
Complex hydrides including the light metal lithium (Li) have 
sufficient gravimetric hydrogen storage capacity, 
and many researches and development about the lithium complex hydride 
such as LiBH$_4$~\cite{c1,c2,c3} and LiNH$_{2}$~\cite{c4,c5,c6} 
have been done recently. 
Particularly, LiBH$_4$ can desorb about $14$~mass\% of hydrogen 
by the following thermal decomposition : 
\begin{equation}
\mbox{LiBH}_4 \rightarrow 
\mbox{LiH} + \mbox{B} + \frac{3}{2} \mbox{H}_2. \label{eq:0} 
\end{equation}
However, the experimental value of enthalpy of this reaction is 
$69$~kJ/mol~H$_2$~\cite{c1} 
and LiBH$_4$ is too stable to release hydrogen at ambient condition. 
High temperature and high pressure are needed 
for rehydriding reaction, so its reversibility becomes a problem 
for practical use. 

Z{\"u}ttel {\it et al.}~\cite{c1} investigated the hydrogen desorption 
from LiBH$_4$ in details, and 
reported that 
the thermal desorption spectra of LiBH$_4$ mixed with SiO$_2$-powder 
exhibited three hydrogen desorption peaks. 
These peaks were observed around 500~K, 550~K and 600~K, 
and the products corresponded to ``LiBH$_{3.6}$'' , ``LiBH$_3$'' and 
``LiBH$_2$'' , respectively. 
The compounds in quotes were nominal compositions 
estimated from amount of desorbed hydrogen. 
This is a strong evidence for the existence of the intermediate compounds , 
and the hydrogen desorption reaction takes place 
in at least two steps not a single step like in Eq.(\ref{eq:0}).
Although low temperature release of hydrogen and improvement of reversibility 
can be expected by use of the intermediate compound, 
little attention has been paid yet. 

In this study, 
the stability of the intermediate compounds of LiBH$_4$ 
has been investigated theoretically and we clarify 
a hydriding/dehydriding process of LiBH$_4$. 
The thermal desorption experiment observed 
a structural transition around 380~K and 
the melting of the compound at 550~K. The stability of the solid phases
at absolute zero temperature is mainly discussed here. 
In Sec.~\ref{sec:meth},  
we describe the details of the computational method 
and the possible intermediate compounds of LiBH$_4$. 
Section~\ref{sec:rslt1} reports 
the calculated results on the stability of some candidate compounds.  
And then, we discuss the hydriding/dehydriding reaction via 
the intermediate compounds. 
Furthermore, 
the electronic structures and $\Gamma $-phonon frequencies 
on the most stable compound are studied. 
We also investigate the stability of the borane complex anions, and then 
compare with the solid compounds. 

\section{Computational details}
\label{sec:meth}
First-principles calculations have been performed 
by the ultrasoft pseudopotential method~\cite{c12} 
based on the density functional theory.~\cite{c13} 
The generalized gradient approximation (GGA) formula~\cite{c14} 
is applied to the exchange-correlation energy. 

The interaction between the ion cores and electrons 
is described by the ultrasoft pseudopotential~\cite{c12}, 
and the norm-conservation constraint~\cite{c140} is imposed on Li 
for the calculation efficiency improvement. 
The scalar-relativistic all-electron calculations are first carried out 
on the free atoms(ions). 
We chose $2s$ and $2p $ states for both Li and B pseudopotential 
as the reference states 
with the cutoff radii of $2$ a.u. (Li) and $1.5$ a.u. (B), respectively. 
A single projector function is used for each angular momentum component. 
The $3d$ state is treated as the local part of pseudopotential. 
The hydrogen pseudopotential is constructed from 
$1s, 2s $ and $2p $ states with the cutoff radii of $1.1$ a.u.($s$) 
and $1.2$ a.u. ($p$). 
We use double projector functions for the $s$ component and 
a single projector function for the $p$ component. 
For all pseudopotentials, 
the pseudo-wave functions and the pseudo-charge-augmentation functions are 
optimized by a method similar to that proposed 
by Rappe {\it et al}.~\cite{c141} 
Also, the partial core correction~\cite{c142} is taken into account for 
Li and B pseudopotentials. 

In the solid-state calculations, the pseudo-wave functions 
are expanded by plane waves with a cutoff energy equal to $15$ hartrees. 
The cutoff energy for the charge density and potential is 
set to be $120$ hartrees. 
The integral over the Brillouin zone is approximated 
by the summation on the {\bf k}-grid elements 
of which the edge lengths closer to the target value of $0.15$\AA$^{-1}$ 
as possible. 
We confirmed that these calculation conditions gave 
a good convergence of energy within 0.002 eV/atom. 
The preconditioned conjugate-gradient technique is employed 
to minimize the Kohn-Sham energy functional. 
A procedure based on the iterative diagonalization scheme~\cite{c144} 
and the Broyden charge mixing method~\cite{c145} is 
adopted in this study. 
Optimization of crystal structures is performed till the atomic forces 
and the macroscopic stresses become less than $5 \times 10^{-4}$ hartree/bohr 
and $0.1$ GPa, respectively. 
During the structural optimization process,
the partial occupation numbers near the Fermi level are determined 
by the Fermi-Dirac distribution function 
with $k_B T = 3 \times 10^{-3}$ hartrees. 
The Helmholtz free-energy functional,~\cite{c15} 
including the entropy term, is minimized 
instead of the Kohn-Sham energy functional. 
Then, the improved tetrahedron method~\cite{c16} is used 
in order to minimize the Kohn-Sham energy functional 
in the optimized structure. 
The dynamical matrix is calculated by the force-constant method~\cite{c17} 
to obtain the $\Gamma $ phonon frequencies. 
The atomic displacement is set to be $0.02$ \AA. 
The further details of calculation are described 
in Refs.~\onlinecite{c2}, \onlinecite{c144} and references therein.

As the candidates of the intermediate compounds, 
the existing alkali metal-B-H materials are used.  
For example, 
it is well known that boron (B) and hydrogen (H) form inorganic compounds 
called 'borane' and 
the compounds of CsB$_3$H$_8$,~\cite{c8} K$_2$B$_6$H$_6$,~\cite{c9} 
and K$_2$B$_{12}$H$_{12}$~\cite{c10} are reported. 
The space group of crystal structure for the compound CsB$_3$H$_8$ is 
$Ama2$ (No.40) and cation Cs$^+$ and anion [B$_3$H$_8$]$^-$ 
are arranged same as the NaCl-type structure with an orthorhombic distortion. 
The crystal structures of K$_2$B$_6$H$_6$ and K$_2$B$_{12}$H$_{12}$ are 
best described as an anti-CaF$_2$-type arrangement with K$^+$ cation 
in the center of a tetrahedron formed by [B$_n$H$_n$]$^{2-}$ dianions. 
We first assumed that LiB$_3$H$_8$, Li$_2$B$_6$H$_6$ and 
Li$_2$B$_{12}$H$_{12}$ compounds, which had the same crystal structures 
as existing CsB$_3$H$_8$, K$_2$B$_6$H$_6$ and K$_2$B$_{12}$H$_{12}$, 
respectively. 
The stability of these candidates was 
evaluated using the first-principles calculations. 
Since the existence of a series of 
{\it closo}-type dianions [B$_n$H$_n$]$^{2-}$($n=5-12$) is 
also well known, 
our calculation have been expanded 
to Li$_2$B$_n$H$_n$ ($n=5-12$) compounds.  
Although Li atom and [B$_n$H$_n$] cluster are arranged 
in the anti-CaF$_2$-type structure 
and the unit cell parameters are supposed to be 
a face-centered-cubic symmetry 
($a=b=c, \alpha = \beta = \delta = 90^\circ$ ) 
as starting points for the structural optimization, 
the output relaxed compounds have different symmetry 
depending on the structure of [B$_n$H$_n$] clusters. 
The alkali metal salt 
of monoclinic K$_2$B$_{12}$(OH)$_{12}$~\cite{c11} with 
{\it closo}-[B$_{12}$(OH)$_{12}$]$^{2-}$ dianions has been reported, 
and monoclinic Li$_2$B$_{12}$H$_{12}$ 
with similar crystal structure is also examined. 

\section{Results and discussions}
\label{sec:rslt1}
\subsection{Stability of candidate compounds : LiB$_x$H$_y$}
\label{sec:s1}
Table~\ref{tab:1} shows the calculated results on 
the structural parameters and the cohesive energies of 
LiB$_3$H$_8$ and Li$_2$B$_n$H$_n$ ($n=5-12$). 
We denote 
the cubic Li$_2$B$_{12}$H$_{12}$ based on K$_2$B$_{12}$H$_{12}$ with type-1 
and monoclinic Li$_2$B$_{12}$H$_{12}$ based on K$_2$B$_{12}$(OH)$_{12}$ 
with type-2, respectively. 
Compared with the cohesive energy of two type of Li$_2$B$_{12}$H$_{12}$, 
the value of monoclinic Li$_2$B$_{12}$H$_{12}$(type-2) 
is larger than that of type-1. 
Therefore, the type-2 Li$_2$B$_{12}$H$_{12}$ is easy to form. 
After the next paragraph, 
the only result concerning monoclinic Li$_2$B$_{12}$H$_{12}$(type-2) 
is shown. 

\begingroup
\squeezetable
\begin{table*}
\caption{\label{tab:1}
Structural parameters and cohesive energies ($E_{coh}$) 
of the candidate compounds : LiB$_x$H$_y$. 
It is denoted that 
the cubic Li$_2$B$_{12}$H$_{12}$ based on K$_2$B$_{12}$H$_{12}$ with type-1 
and monoclinic Li$_2$B$_{12}$H$_{12}$ based on K$_2$B$_{12}$(OH)$_{12}$ 
with type-2, respectively. 
}
\begin{ruledtabular}  
\begin{tabular}{ccccccccc}
Compound & Space group & Unit cell parameters & 
\multicolumn{5}{c}{Atomic position} & $E_{coh}$ \\
 & & & Atom & Site & $x/a$ & $y/b$ & $z/c$ & (eV/atom) \\
\hline
LiB$_3$H$_8$ & $Ama2 $ & $a=9.188$ \AA & 
Li   &  $4b$ & $0.0595$ & $0.6321$ & $0.25$ & $3.3770$ \\ 
 & $(No.40)$ & $b=8.813$ \AA & 
B1   &  $4b$ & $0.1372$ & $0.1335$ & $0.25$ & \\ 
 & & $c=5.763$ \AA & 
B2   &  $8c$ & $0.3090$ & $0.1331$ & $0.0945$ & \\ 
 & & & H1   &  $4b$ & $0.0690$ & $0.2507$ & $0.25$ & \\ 
 & & & H2   &  $4b$ & $0.0684$ & $0.1666$ & $0.25$ & \\ 
 & & & H3   &  $8c$ & $0.1386$ & $0.5149$ & $0.5190$ & \\ 
 & & & H4   &  $8c$ & $0.1823$ & $0.1333$ & $0.0063$ & \\ 
 & & & H5   &  $8c$ & $0.3619$ & $0.2508$ & $0.0186$ & \\
Li$_2$B$_5$H$_5$ & $R3m $ & $a=5.599$ \AA & 
  Li1 & $3a$ & $0$ & $0$ & $0.2755$ & $3.7910$ \\ 
 & $(No.160)$ & $c=16.763$ \AA & 
  Li2 &  $3a$ & $0$ & $0$ & $0.7513$ & \\ 
 & (hexagonal axes) & & B1 &  $3a$ & $0$ & $0$ & $0.0876$ & \\ 
 & & & B2 &  $3a$ & $0$ & $0$ & $0.9295$ & \\ 
 & & & B3 &  $9b$ & $0.1070$ & $0.8930$ & $0.0084$ & \\ 
 & & & H1 &  $3a$ & $0$ & $0$ & $0.1595$ & \\ 
 & & & H2 &  $3a$ & $0$ & $0$ & $0.8586$ & \\ 
 & & & H3 &  $9b$ & $0.2270$ & $0.7730$ & $-0.0078$ & \\ 
Li$_2$B$_6$H$_6$ & $Fm\bar{3}m $ & $a=7.968$ \AA & 
  Li   &  $8c$ & $0.25$ & $0.25$ & $0.25$ & $4.0429$ \\ 
  & $(No.225)$ & & B  &  $24e$ & $0.1553$ & $0$ & $0$ & \\ 
  & & & H  &  $24e$ & $0.3054$ & $0$ & $0$ & \\ 
Li$_2$B$_7$H$_7$ & $I2 $ & $a=5.691$ \AA & 
  Li1   &  $2b$ & $0$ & $0.1501$ & $0.5$ & $4.0850$ \\ 
 & $(No.5)$ & $b=9.836$ \AA & Li2 & $2b$ & $0.5$ & $0.2317$ & $0$ & \\
 & & $c=5.625$ \AA & B1 & $2a$ & $0$ & $0.1564$ & $0$ & \\
 & & $\beta =77.39^\circ $ & B2 & $4c$ & $-0.1871$ & $0.0529$ & $-0.1063$ & \\
 & & & B3 & $4c$ & $0.1160$ & $-0.1087$ & $0.0659$ & \\
 & & & B4 & $4c$ & $-0.1335$ & $0.0087$ & $0.1903$ & \\
 & & & H1 & $2a$ & $0$ & $0.2792$ & $0$ & \\
 & & & H2 & $4c$ & $-0.3463$ & $0.0805$ & $-0.2099$ & \\
 & & & H3 & $4c$ & $0.2215$ & $-0.2030$ & $0.1343$ & \\
 & & & H4 & $4c$ & $-0.2700$ & $0.0109$ & $0.3879$ & \\
Li$_2$B$_8$H$_8$ & $I\bar{4}2m $ & $a=5.572 $ \AA & 
 Li & $4d$ & $0$ & $0.5$ & $0.25$ & $4.0920$ \\ 
 & $(No.121)$ & $c=10.687$ \AA &  B1 & $8i$ & $0.1032$ & $0.1032$ & $-0.1242$ &  \\ 
 & & &  B2 & $8i$ & $0.1631$ & $0.1631$ & $0.0304$ &  \\ 
 & & &  H1 & $8i$ & $0.2031$ & $0.2031$ & $-0.2112$ &  \\ 
 & & &  H2 & $8i$ & $0.3114$ & $0.3114$ & $0.0580$ &  \\ 
Li$_2$B$_9$H$_9$ & $R3 $ & $a=7.044$ \AA & 
Li1 & $3a$ & $0$ & $0$ & $0.2482$ & $4.1326$ \\ 
 & $(No.146)$ & $c=15.062$ \AA & Li2 & $3a$ & $0$ & $0$ & $0.8070$ &  \\ 
 & (hexagonal axes)& & B1 & $9b$ & $0.8087$ & $0.9031$ & $0.9508$ &  \\ 
 & & & B2 & $9b$ & $0.8118$ & $0.9047$ & $0.0694$ &  \\ 
 & & & B3 & $9b$ & $0.8661$ & $0.7287$ & $0.0096$ &  \\ 
 & & & H1 & $9b$ & $0.6730$ & $0.8346$ & $0.8923$ &  \\ 
 & & & H2 & $9b$ & $0.6756$ & $0.8357$ & $0.1281$ &  \\ 
 & & & H3 & $9b$ & $0.7699$ & $0.5322$ & $0.0141$ &  \\ 
Li$_2$B$_{10}$H$_{10}$ & $I422 $ & $a=6.196$ \AA & 
 Li & $4d$ & $0$ & $0.5$ & $0.25$ & $4.2360$ \\ 
 & $(No.97)$ & $c=10.356$ \AA &  B1 & $4e$ & $0$ & $0$ & $0.1827$ &  \\ 
 & & & B2 & $16k$ & $0.1942$ & $-0.0809$ & $0.0739$ &  \\ 
 & & & H1 & $4e$ & $0$ & $0$ & $0.2984$ &  \\ 
 & & & H2 & $16k$ & $0.3647$ & $-0.1582$ & $0.1070$ &  \\ 
Li$_2$B$_{11}$H$_{11}$ & $I2 $ & $a=7.083$ \AA & 
 Li1 & $2b$ & $0$ & $-0.1091$ & $0.5$ & $4.2397$ \\ 
 & $(No.5)$ & $b=11.028$ \AA & Li2 & $2b$ & $0.5$ & $-0.1083$ & $0$ & \\ 
 & & $c=7.080$ \AA & B1 & $2a$ & $0$ & $-0.1264$ & $0$ & \\ 
 & & $\beta = 70.52^\circ$ & B2 & $4c$ & $0.1117$ & $0.1190$ & $-0.1120$ & \\ 
 & & & B3 & $4c$ & $0.0835$ & $-0.0162$ & $-0.2427$ &  \\ 
 & & & B4 & $4c$ & $0.2425$ & $-0.0162$ & $-0.0840$ &  \\ 
 & & & B5 & $4c$ & $0.1293$ & $0.0825$ & $0.1290$ &  \\ 
 & & & B6 & $4c$ & $0.1448$ & $-0.0795$ & $0.1446$ &  \\ 
 & & & H1 & $2a$ & $0$ & $-0.2355$ & $0$ &  \\ 
 & & & H2 & $4c$ & $0.2016$ & $0.2052$ & $-0.2020$ &  \\ 
 & & & H3 & $4c$ & $0.4160$ & $-0.0284$ & $-0.1910$ &  \\ 
 & & & H4 & $4c$ & $0.1902$ & $-0.0283$ & $-0.4165$ &  \\ 
 & & & H5 & $4c$ & $0.2133$ & $0.1471$ & $0.2127$ &  \\ 
 & & & H6 & $4c$ & $0.2305$ & $-0.1441$ & $0.2300$ &  \\ 
Li$_2$B$_{12}$H$_{12}$  & $Fm\bar{3} $ & $a=10.083$ \AA &
 Li & $8c$ & $0.25$ & $0.25$ & $0.25$ & $4.2997$ \\ 
(type-1) & $(No.202)$ & & B & $48h$ & $0$ & $0.1448$ & $0.0881$ & \\
 & & & H & $48h$ & $0$ & $0.2490$ & $0.1469$ & \\
Li$_2$B$_{12}$H$_{12}$  & $P2_1/n $ & $a=7.358$ \AA & 
Li    & $4e$ & $0.6747$ & $0.6250$ & $0.5323$ & $4.3461$ \\ 
(type-2) & $(No.14)$ & $b=9.556$ \AA & 
B1    & $4e$ & $0.6770$ & $0.5392$ & $0.1632$ & \\
 & & $c=6.768$ \AA & 
B2    & $4e$ & $0.4447$ & $0.5629$ & $0.2239$ & \\
 & & $\beta = 92.26^\circ$ & 
B3    & $4e$ & $0.5425$ & $0.6760$ & $0.0442$ & \\
 & & & B4    & $4e$ & $0.6915$ & $0.5689$ & $0.9022$ & \\
 & & & B5    & $4e$ & $0.6856$ & $0.3922$ & $0.9959$ & \\
 & & & B6    & $4e$ & $0.5329$ & $0.3887$ & $0.1933$ & \\
 & & & H1    & $4e$ & $0.8033$ & $0.5642$ & $0.2786$ & \\
 & & & H2    & $4e$ & $0.4008$ & $0.6050$ & $0.3840$ & \\
 & & & H3    & $4e$ & $0.5707$ & $0.7983$ & $0.0742$ & \\
 & & & H4    & $4e$ & $0.8252$ & $0.6151$ & $0.8271$ & \\
 & & & H5    & $4e$ & $0.8231$ & $0.3231$ & $0.9941$ & \\
 & & & H6    & $4e$ & $0.5558$ & $0.3097$ & $0.3310$ & 

\end{tabular}  
\end{ruledtabular}  
\end{table*}
\endgroup

The enthalpy of formation for hydriding reactions 
including LiBH$_4$ from LiH (NaCl-type),  
$\alpha$-B (rhombohedral), 
and H$_2$ molecule are given in Table~\ref{tab:2}, 
where the zero-point energy corrections are not taken into consideration.  
They are provided 
using calculated cohesive energies of $2.3609$~eV/atom for LiH, 
$6.2013$~eV/atom for $\alpha$-B, $2.2689$~eV/atom for H$_2$ molecule, 
and $3.1501$~eV/atom for LiBH$_4$(orthorhombic $Pnma$ symmetry), respectively. 
The enthalpies of formation for Li$_2$B$_n$H$_n$ ($n=10,11,12$) 
are more negative than that for LiBH$_4$. 
Therefore these compounds have a great potential for 
generating in hydriding reactions from LiH and $\alpha$-B 
as the intermediate phase of LiBH$_4$. 

\begin{table}
\caption{\label{tab:2}
The enthalpies of formation for the hydriding reaction 
of various compounds LiB$_x$H$_y$ including LiBH$_4$ from LiH, $\alpha$-B, 
and H$_2$ molecule, where the zero-point energy corrections 
are not taken into consideration. }
\begin{ruledtabular}
\begin{tabular}{cc}
Hydriding reaction & $\begin{array}{c} \mbox{Enthalpy of formation} \\
                                       \mbox{(kJ/mol~H$_2$)} \end{array}$ \\
\hline
 LiH +  3B + 7/2H$_2$ $\rightarrow$ LiB$_3$H$_8$ & $-36$ \\
2LiH +  5B + 3/2H$_2$ $\rightarrow$ Li$_2$B$_5$H$_5$ & $113$ \\
2LiH +  6B +  2 H$_2$ $\rightarrow$ Li$_2$B$_6$H$_6$ & $-42$ \\
2LiH +  7B + 5/2H$_2$ $\rightarrow$ Li$_2$B$_7$H$_7$ & $-45$ \\
2LiH +  8B +  3 H$_2$ $\rightarrow$ Li$_2$B$_8$H$_8$ & $-32$ \\
2LiH +  9B + 7/2H$_2$ $\rightarrow$ Li$_2$B$_9$H$_9$ & $-42$ \\
2LiH + 10B +  4 H$_2$ $\rightarrow$ Li$_2$B$_{10}$H$_{10}$ & $-87$ \\
2LiH + 11B + 9/2H$_2$ $\rightarrow$ Li$_2$B$_{11}$H$_{11}$ & $-79$ \\
2LiH + 12B +  5 H$_2$ $\rightarrow$ Li$_2$B$_{12}$H$_{12}$ & $-125$ \\
\hline
 LiH +   B + 3/2H$_2$ $\rightarrow$ LiBH$_4$ & $-75$
\end{tabular}  
\end{ruledtabular}  

\end{table}

\begin{figure}
\includegraphics[width=9cm]{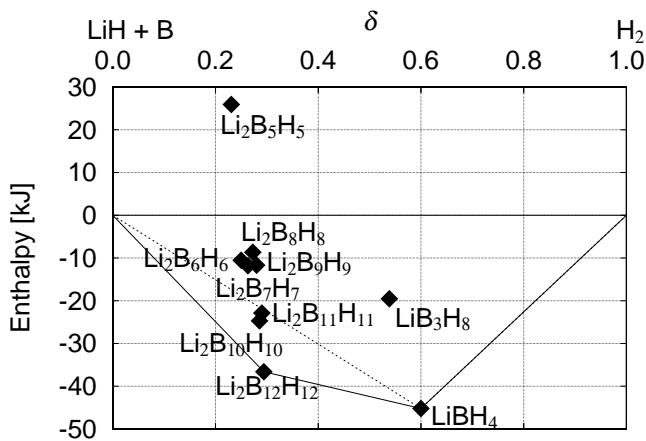}

\vspace{-0.3cm}
\caption{\label{fig:1}
Enthalpy of formation 
relative to $(1-\delta )$[LiH + B] and molecular H$_2$ of $\delta $ mole 
(reaction Eq.(\ref{eq:1}) in text), 
as a function of the mole fraction of H$_2$ , $\delta$.}
\end{figure}

In order to understand the stability of each compound intuitively, 
we represent the enthalpy of formation 
corresponding to the following reaction 
with hydrogen of $\delta $ mole in Fig.~\ref{fig:1} : 
\begin{equation} 
(1-\delta )[ \mbox{LiH} + \mbox{B}] + \delta \mbox{H}_2 \rightarrow 
\frac{1-\delta }{x} \mbox{LiB}_x \mbox{H}_y 
+ \frac{(1-\delta )(x-1)}{x} 
{\rm LiH} , \label{eq:1}
\end{equation} 
where $\delta = (y-1)/(2x+y-1)$ shows the mole fraction of H$_2$ and 
$x \ge 1$. 
From Fig.~\ref{fig:1}, 
monoclinic Li$_2$B$_{12}$H$_{12}$ is formed 
as the intermediate compound of 
the hydriding/dehydriding reaction of LiBH$_4$, 
because the actual reaction goes along the lowest state of enthalpy. 
Therefore, following hydriding/dehydriding process 
is proposed : 
\begin{eqnarray}
\mbox{LiBH}_4 &\leftrightarrow & 
\frac{1}{12}\mbox{Li}_2\mbox{B}_{12}\mbox{H}_{12}
 + \frac{5}{6}\mbox{LiH} + \frac{13}{12}\mbox{H}_2 , \label{eq:2} \\
 &\leftrightarrow & \mbox{LiH} + \mbox{B} + \frac{3}{2}\mbox{H}_2. \label{eq:3}
\end{eqnarray} 
The enthalpy without zero-point energy effects 
and hydrogen content of the first reaction (Eq.(\ref{eq:2})) 
are $56$~kJ/mol~H$_2$ and $10$~mass\% 
and those of the second reaction (Eq.(\ref{eq:3})) are $125$~kJ/mol~H$_2$ 
and $4$~mass\%, respectively. 
These results agree with the experimental ones, 
which reported that $9$~mass\% of hydrogen are liberated 
in the hydrogen desorption peak 
in thermal desorption spectra of LiBH$_4$ 
mixed with SiO$_2$-powder in Ref.~\onlinecite{c1}. 
The enthalpy value of $56$~kJ/mol~H$_2$ for the first reaction 
is lower than computational one of $75$~kJ/mol~H$_2$ 
for direct dehydriding reaction of LiBH$_4$ (Eq.(\ref{eq:0})).~\cite{c2}  
So the hydrogen desorption and absorption can occur 
at low temperature and pressure comparatively 
by using this intermediate compound. 

We also calculated other possible crystal structures 
such as Cu$_2$B$_{10}$H$_{10}$-type~\cite{cubh} and 
[Li(thp)$_3$]$_2$[B$_{11}$H$_{11}$]-type.~\cite{libh11} 
Moreover, we performed the calculation for 
the $\Gamma $-phonon frequencies of the candidate materials. 
If there was a soft-mode, 
we lowered crystal structure symmetry 
with moving the atoms along the direction of the soft-mode 
eigenvectors.  
The maximum gain of cohesive energy was $0.03$~eV/atom 
for Li$_2$B$_6$H$_6$, 
however, no compound or crystal structure more stable 
than monoclinic Li$_2$B$_{12}$H$_{12}$ was found. 
The present calculation result on the energy of 
monoclinic Li$_2$B$_{12}$H$_{12}$ provides the upper limit value 
for the thermodynamic stability. 
In conclusion, 
the existence of the intermediate compound of LiBH$_4$  
was predicted theoretically. 

Recently Kang {\it et al.}~\cite{apl1} have reported that LiBH and LiB are 
the intermediate phases of LiBH$_4$, and propose the dehydriding reaction 
of LiBH$_4$ through LiBH as follows : 
\begin{equation}
\mbox{LiBH}_4 \rightarrow \mbox{LiBH} + \frac{3}{2}\mbox{H}_2 . \label{eq:4}
\end{equation}
We also performed the first-principles calculation of 
the orthorhombic phase of {\it Pnma} LiBH. 
The reaction enthalpy for Eq.(\ref{eq:4}) is 
$1.30$eV per LiBH$_4$ formula unit (84kJ/molH$_2$), 
which is in good agreement with reported value 
of $1.28$eV per LiBH$_4$ formula unit in Ref.\onlinecite{apl1}. 
On the other hand, the reaction enthalpy of Eq.(\ref{eq:2}) 
is $0.63$eV per LiBH$_4$ formula unit. 
Therefore, the dehydriding reaction via the intermediate compound 
Li$_2$B$_{12}$H$_{12}$ is energetically more preferable one.

\subsection{Intermediate compound : Li$_2$B$_{12}$H$_{12}$}
Here we describe the fundamental properties, such as 
the crystal structure, the electronic structure, 
and the $\Gamma $ phonon frequencies, on Li$_2$B$_{12}$H$_{12}$ which is 
expected as intermediate compound of LiBH$_4$. 

\begin{figure}
\includegraphics[width=8cm]{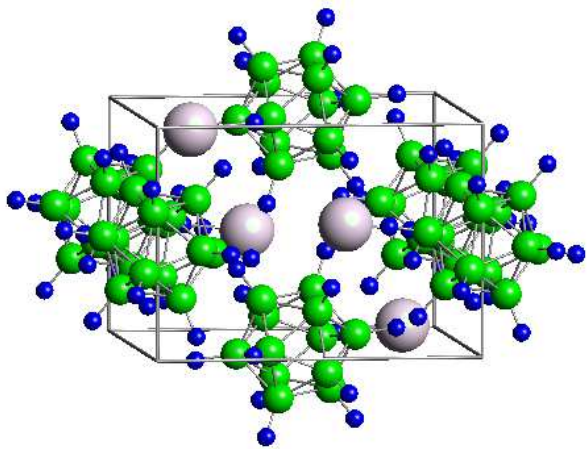}

\vspace{-0.3cm}
\caption{\label{fig:3}
(Color online) Crystal structure model of  
monoclinic Li$_2$B$_{12}$H$_{12}$(type-2). 
Large, middle, and small spheres denote Li, B, and H atoms, respectively. }
\end{figure}

Optimized crystal structure model 
of monoclinic Li$_2$B$_{12}$H$_{12}$ is shown in 
Fig.~\ref{fig:3}. 
The bond lengths between B and H atom of Li$_2$B$_{12}$H$_{12}$
are $1.20-1.21$ \AA , 
and they are very close to those for LiBH$_4$ ($1.23-1.24$ \AA) 
reported in Ref.~\onlinecite{c2}. 
As for the B-B bond lengths of intra-icosahedron, 
the values of $1.779 - 1.811$~\AA \  for Li$_2$B$_{12}$H$_{12}$ 
are comparable to the experimental ones 
for the $\alpha$-rhombohedral boron ($\alpha $-B)
of $1.751-1.806$ \AA,~\cite{expb} too. 
Since boron crystal has the icosahedral B$_{12}$ cluster 
as a common structural component, 
the decomposition of [B$_{12}$H$_{12}$]$^{2-}$ anion into 
B$_{12}$ cluster and hydrogen molecule is easy to understand.  

\begin{figure} 
\includegraphics[width=9cm]{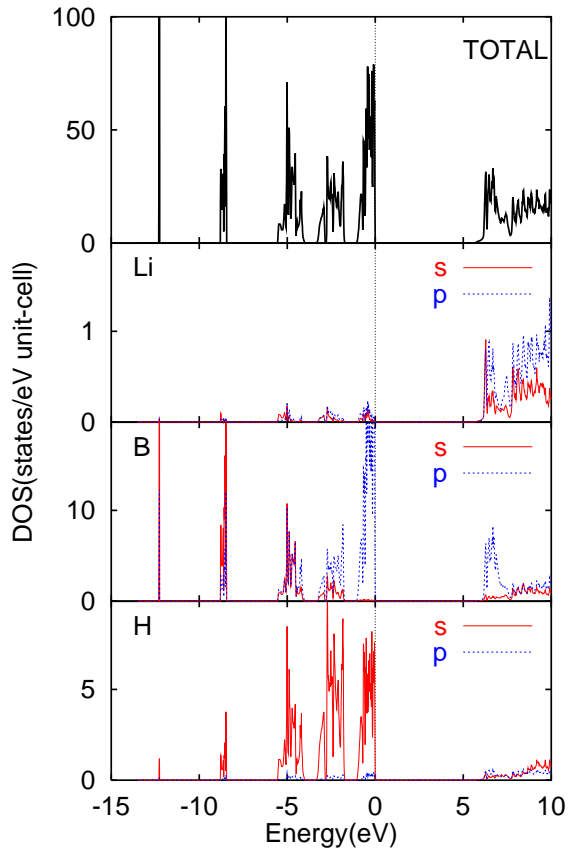} 
 
\vspace{-0.3cm} 
\caption{\label{fig:4} 
(Color online) 
The total and partial densities of states (DOS) for Li$_2$B$_{12}$H$_{12}$. 
The energy is measured in eV relative to the top of valence states.}
\end{figure}

Figure~\ref{fig:4} shows 
the total and partial density of states (DOS) for Li$_2$B$_{12}$H$_{12}$, 
which denotes that it has the energy gap of $5.60$~eV. 
Since there is little contribution of Li orbital for occupied states, 
Li$_2$B$_{12}$H$_{12}$ consists of Li$^{+}$ and [B$_{12}$H$_{12}]^{2-}$ ions. 
The orbitals of B and H hybridize each other and 
the feature of occupied states in DOS is analogous 
with the distribution of the computed energies 
of the bonding molecular orbitals 
in [B$_{12}$H$_{12}$]$^{2-}$.~\cite{King}

\begin{figure}
\includegraphics[width=9cm]{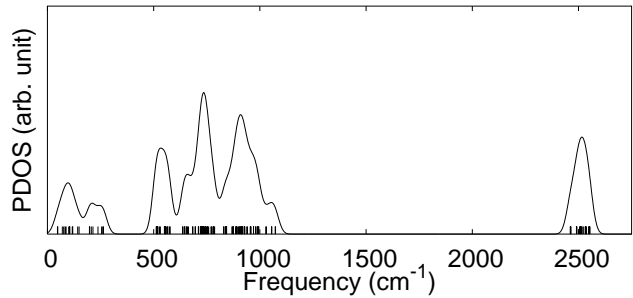}

\vspace{-0.3cm}
\caption{\label{fig:5}
Phonon density of states for monoclinic Li$_2$B$_{12}$H$_{12}$. 
The contribution of the TO $\Gamma $-phonon modes indicated 
by vertical bars is only taken into account 
and the Gaussian broadening with a width of $30$ cm$^{-1}$ is used.}
\end{figure}

The $\Gamma$-phonon mode frequencies of monoclinic Li$_2$B$_{12}$H$_{12}$ 
have been calculated so that the vibrational properties can be compared with 
experiments easily and directly. 
The phonon density of states is shown in Fig.~\ref{fig:5}. 
It is divided into three regions in the same case as LiBH$_4$. 
The first region is less than $300$~cm$^{-1}$ 
where the displacements of Li atoms are dominant. 
The second region is between $450$~cm$^{-1}$ and $1100$~cm$^{-1}$ 
where the B-H bond of [B$_{12}$H$_{12}$]$^{2-}$ dianions vibrates 
with changing in angle between them (bending modes), 
and the third region is between $2400$~cm$^{-1}$ and $2600$~cm$^{-1}$ 
where the inter-atomic distance of B-H bond is changing 
along bond axis (stretching modes). 
The frequencies of bending modes are lower than LiBH$_4$ 
described in Ref.~\onlinecite{c2}, while 
stretching modes are higher. 

The investigation using in situ Raman spectroscopy is effective 
for the confirmation of the short-range order 
or bonding of LiBH$_4$.~\cite{c19} 
We examine the atomistic vibrations of LiBH$_4$ 
during heating 
by in situ Raman spectroscopy, 
and the identification of spectra modes 
originating from Li$_2$B$_{12}$H$_{12}$ is now in progress.~\cite{aplo}  

\subsection{Stability of complex anions : [B$_n$H$_n$]$^{2-}$}
Finally, we consider the relation between the stability of the 
compounds ${\rm Li_2B}_n{\rm H}_n$ and those of complex anions 
$[{\rm B}_n{\rm H}_n]^{2-}$. 
The energies of the isolated $[{\rm B}_n{\rm H}_n]^{2-}$ are obtained 
using a face-centered-cubic supercell 
with $a=20$~{\AA}. The single $\Gamma$ point 
is used for the {\bf k}-point sampling. 
The energies for the charged systems 
are computed by adding uniform background charges and 
improved with Makov and Payne correction~\cite{Makov} 
for the interaction between periodic image charges.

\begin{figure}
\begin{center}
\includegraphics[width=9cm]{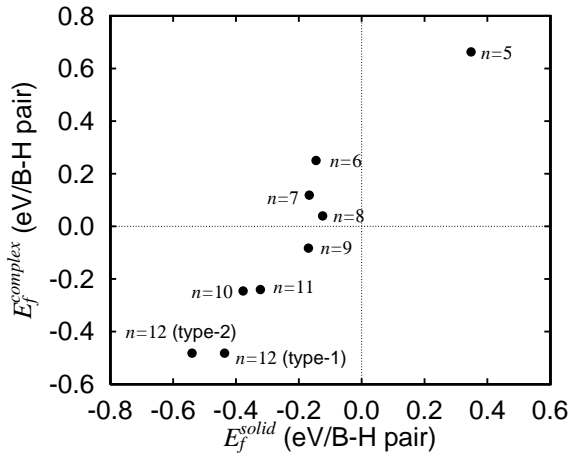} 

\end{center}
\caption{\label{fig:BH}
Comparison of the formation energies, 
$E_f^{solid}$ and $E_f^{complex}$, corresponding to the 
reactions Eq.(\ref{eq:Esolid}) and Eq.(\ref{eq:Ecmplx}) in text, respectively, 
where the energies are normalized by the number of B-H pairs. }
\end{figure}

Figure~\ref{fig:BH} shows the comparison of the formation energies, 
$E_f^{solid}$ and $E_f^{complex}$, corresponding to the following 
reactions:
\begin{equation}
 2{\rm LiH} + n{\rm B} + \frac{n-2}{2} {\rm H_2} \leftrightarrow
 {\rm Li}_2{\rm B}_n{\rm H}_n, \label{eq:Esolid}
\end{equation}
and
\begin{equation}
 n{\rm B} + \frac{n}{2} {\rm H_2} + 2e^- \leftrightarrow
 [{\rm B}_n{\rm H}_n]^{2-}, \label{eq:Ecmplx}
\end{equation}
where the energies are normalized by the number of B-H pairs for 
comparison purposes. 
We can find a fairly good correlation between both energies. 
This is probably due to the fact that the electrostatic interaction 
between ${\rm Li}^+$ and $[{\rm B}_n{\rm H}_n]^{2-}$ is not sensitive 
to $n$ and a large part of this energy cancels out with that of LiH 
in Eq.~(\ref{eq:Esolid}). 
Among the {\it closo}-type dianions considered here, 
$[{\rm B}_{12}{\rm H}_{12}]^{2-}$ is the most stable one. 

${\rm LiBH_4}$ desorbs hydrogen at  
temperatures above the melting point. 
The latent heat of fusion has been reported~\cite{Smith} to be  
0.078~eV/formula-unit for ${\rm LiBH_4}$ and similar values are 
expected for ${\rm Li}_2{\rm B}_n{\rm H}_n$, which is considerably 
smaller than the energy difference between 
$[{\rm B}_{12}{\rm H}_{12}]^{2-}$ and other {\it close}-type dianions. 
The stability of the $[{\rm B}_{12}{\rm H}_{12}]^{2-}$ anion supports 
our main conclusion in this study, that is, the intermediate phase 
suggested by the experiment is  ${\rm Li}_2{\rm B}_{12}{\rm H}_{12}$. 

\section{Summary}
\label{sec:conc}
We have examined 
the stability of LiB$_3$H$_8$ and 
Li$_2$B$_n$H$_n (n=5-12)$ , 
which are the possible intermediate compounds of LiBH$_4$, 
by the first-principles calculation. 
Our computational results for enthalpy of the hydriding reactions 
provide that monoclinic Li$_2$B$_{12}$H$_{12}$ 
is the most stable one among the candidates. 
The following hydriding/dehydriding process 
of LiBH$_4$ is proposed : 
LiBH$_4 \leftrightarrow 1/12$Li$_2$B$_{12}$H$_{12} + 5/6$LiH $+ 13/12$H$_2
\leftrightarrow $ LiH $+$ B $+$ $3/2$H$_2$. 
The hydrogen content of the first and the second reaction are 
$10$~mass\% and $4$~mass\%, respectively, 
which agree well with the thermal desorption spectra (TDS) experiment 
on LiBH$_4$.~\cite{c1} 
The heat of formation without zero point energy corrections 
for the first reaction, which is estimated  
from the solid state LiBH$_4$ with $Pnma$ symmetry (not liquid LiBH$_4$),  
is $56$~kJ/molH$_2$. 
This value is lower than that 
for the direct reaction 
(LiBH$_4 \leftrightarrow $ LiH $+$ B $+ \frac{3}{2}$H$_2$). 
Therefore,  
low temperature release of hydrogen 
can be expected by use of this intermediate compound. 

We have calculated the electronic structure and 
the $\Gamma$-phonon frequencies 
of monoclinic Li$_2$B$_{12}$H$_{12}$. 
This compound has the energy gap of $5.60$~eV and 
consists of Li$^{+}$ and [B$_{12}$H$_{12}]^{2-}$ ions. 
From the phonon density of states, 
it is predicted that its bending modes have lower frequencies 
than that of LiBH$_4$,~\cite{c2} while stretching modes are higher. 
The identification of the experimental Raman spectra modes 
originating from Li$_2$B$_{12}$H$_{12}$ 
is now in progress. 

The stability of {\it closo} borane complex anions [B$_n$H$_n$]$^{2-}$
was also examined. 
We found a fairly good correlation between 
the formation energies of the solid phases 
and the isolated dianions. 
This result supports the validity of the intermediate compound 
indicated in the TDS experiment 
being ${\rm Li}_2{\rm B}_{12}{\rm H}_{12}$. 

There are various kinds of borane,  
such as {\it nido}-type [B$_n$H$_{n+4}$]
and {\it arachno}-type [B$_n$H$_{n+6}$], 
in addition to 
{\it closo}-type borane [B$_n$H$_{n+2}$].~\cite{c7} 
The {\it nido} and {\it arachno} borane are derived from 
{\it closo} borane by removing one and two vertices, respectively. 
These generate the salts in the ionized state 
as well as {\it closo}-type dianions. 
These alkali metal salts are also 
the candidates of the intermediate compound of LiBH$_4$. 
We will study the details of the hydriding/dehydriding process for LiBH$_4$ 
including the stability of these materials in future. 

\begin{acknowledgments}
We wish to thank M. Matsumoto, R. Jinnouchi and S. Hyodo 
for helpful discussions. 
This work was partially supported by the New Energy and Industrial Technology 
Development Organization(NEDO), International Joint Research 
under the "Development for Safe Utilization and Infrastructure of Hydrogen" 
Project (2004-2005).
\end{acknowledgments}





\end{document}